\documentstyle[12pt]{article}
\begin{document}

\baselineskip 24pt

\newcommand{\sheptitle}
{Supersymmetric Higgs Bosons \\ at the Limit}

\newcommand{\shepauthor}
{T. Elliott, S. F. King and P. L. White,}

\newcommand{\shepaddress}
{Physics Department,\\University of Southampton,\\Southampton,
\\SO9 5NH,\\U.K.}

\newcommand{\shepabstract}
{We obtain an improved upper bound on the
lightest neutral CP-even
Higgs boson mass from a low energy renormalisation-group
analysis of the Higgs sector of the
next-to-minimal supersymmetric standard model.
We find $m_h< 145$ GeV for $m_t=90$ GeV,
decreasing to $m_h< 123$ GeV for $m_t=180$ GeV.
We also discuss the light Higgs spectrum in the
region of parameter space close to the upper bound.}

\begin{titlepage}
\hfill SHEP 92/93-11

\hfill hep-ph/9302202
\vspace{.4in}
\begin{center}
{\Huge{\bf \sheptitle}}
\bigskip \\ \shepauthor \\ {\it \shepaddress} \\ \vspace{.5in}
{\bf Abstract} \bigskip \end{center} \setcounter{page}{0}
\shepabstract
\end{titlepage}

The question of the origin of electroweak mass is one
of the most urgent questions of present day particle physics.
The discovery of a particle which resembles the Higgs boson
of the minimal standard model, and the measurement of its mass,
will provide clues as to the nature of new physics beyond the
standard model.
In this letter we shall be concerned with the question
of how heavy the lightest neutral CP-even
supersymmetric Higgs boson,
$h^0$, can be within the framework of supersymmetric grand
unified theories
(SUSY GUTs) \cite{1}.
In SUSY GUTs all the Yukawa couplings are constrained
to remain perturbative in the region
$M_{SUSY}\sim 1$ TeV to $M_{GUT}\sim 10^{16}$ GeV.
This constraint provides a maximum value at low energies for those
Yukawa couplings which are not asymptotically-free, and is
obtained from the renormalisation group (RG) equations
together with the boundary
conditions that the couplings become non-perturbative
at $M_{GUT}$ -- the so-called ``triviality limit''.
In the minimal supersymmetric standard model (MSSM)
\cite{2}, the triviality limits provide a useful bound on the
top quark mass
$m_t$. The upper bound on the $h^0$ mass, $m_h$, in the MSSM,
including radiative corrections,
has recently been the subject of much discussion\cite{6,7,8,9,10,11,12}.
However the MSSM is not the most general
low energy manifestation of SUSY \hbox{GUTs.}
It is possible that SUSY GUTs give
rise to a low energy theory which contains an additional
gauge singlet field, the so called next-to-minimal supersymmetric
standard model (NMSSM) \cite{3,4,5}.
Here we shall concentrate on the question
of the upper-bound on $m_h$ in the NMSSM
which is obtained from triviality limits of Yukawa couplings.

In the NMSSM there are two Higgs doublets $H_1,H_2$ and a
complex Higgs singlet $N$,
leading to three neutral CP-even Higgs
bosons, two
neutral CP-odd states and two charged bosons
in the physical spectrum.
The superpotential has the form
\begin{equation}
W=h_tQt^cH_2+ \lambda NH_1H_2 -  \frac{k}{3}N^3 + \ldots,
\end{equation}
where the superfield Q contains the left-handed
third family quark doublet $(t_L, b_L)$,
the superfield $t^c$ contains the
charge conjugate of the right-handed top quark field,
$H_{1,2}$ contain the Higgs doublets and $N$ contains the Higgs
gauge singlet. The ellipsis represents terms whose relatively small
couplings will not play a role in our analysis.
The NMSSM removes the $\mu H_1H_2$ term in the superpotential
of the MSSM and replaces its effect by the
vacuum expectation value (VEV) of the Higgs singlet, $<N>=x$.
The other VEVs are $<H_{1,2}>=v_{1,2}$,
where $v=\sqrt{{v_1}^2 + {v_2}^2}=174$ GeV.
The trilinear term $\frac{k}{3}N^3$ is necessary in order
to avoid an approximate global $U(1)$ symmetry
(broken by instanton effects)
which, when spontaneously broken by the VEVs, would lead to an axion.
The Higgs sector of the NMSSM
reduces to that of the MSSM in the limit
$r=x/ v\rightarrow \infty$ with $\lambda x$ and $kx$ fixed,
where one of the neutral CP-even states and one of the
CP-odd states decouple \cite{4}.

An upper bound on the lightest neutral CP-even scalar $h^0$
in the NMSSM
may be obtained from the real symmetric
$3\times 3$ neutral scalar mass squared matrix,
by using the fact that the smallest eigenvalue of such a matrix
must be smaller than the smallest eigenvalue
of its upper $2\times 2$ block.
The resulting bound at tree-level is \cite{5}
\begin{equation}
{m_{h}}^2\leq {M_Z}^2 +
(\lambda ^2v^2-{M_Z}^2)\sin^22\beta.
\end{equation}
where $\tan \beta \equiv \frac{v_2}{v_1}$,
and $\lambda$ is regarded as a running parameter
evaluated at $M_{SUSY}$.
The upper bound on $m_h$ is determined by
the maximum value of $\lambda (M_{SUSY})$, henceforth
denoted $\lambda _{max}$.
The value of $\lambda _{max}$ is obtained
by solving the SUSY RG equations for the Yukawa couplings
$h_t$, $\lambda$
and $k$ in the region
$M_{SUSY}=1$ TeV to $M_{GUT}=10^{16}$ GeV \cite{13,14}.
If the Yukawa couplings $h_t$, $\lambda$ and $k$
are all initially large at $M_{GUT}$ then they approach low energy
fixed point ratios \cite{13}. However,
if the boundary condition at $M_{GUT}$ is $\lambda \gg k$,
then larger values of $\lambda (M_{SUSY})$ can be achieved.
We have repeated the calculation
of ref.\cite{14} and found that
for $h_t(M_{SUSY})=0.5-1.0$,
${\lambda}_{max}=0.87-0.70$ and
for $h_t(M_{SUSY}) \rightarrow 1.06$,
\footnote{ $h_t(M_{SUSY})\leq 1.06$ is
the triviality bound, which, together with
$m_t=h_t(m_t)v\sin \beta$, where $h_t(m_t)\leq 1.12$,
implies the bound $m_t \leq 195$ GeV.}
${\lambda}_{max}\rightarrow 0$ (with $k=0$ always).
Radiative corrections to the tree-level bound
in Eq.(2) have been
considered in refs.\cite{15,16,17}.
In ref.\cite{16} these
were estimated from a low energy RG analysis of the Higgs sector
of the model between $M_{SUSY}$ and a lower scale $\mu$,
assuming that only one Higgs boson has a mass below
$M_{SUSY}$. Here we shall consider
the more general case in which both Higgs doublets and the
Higgs singlet may
be lighter than $M_{SUSY}$, making the usual approximation
of hard decoupling below $M_{SUSY}$ of the superpartners.
We shall then apply this technique to obtain a new upper bound on $m_h$,
and supplement the discussion
with a full numerical analysis
of the light Higgs spectrum of the NMSSM in the region
of parameter space close to this upper bound.

At energy scales above $M_{SUSY}$ the Higgs potential in the NMSSM
with general soft SUSY breaking terms is given from Eq.(1) by
\begin{eqnarray}
V_{Higgs} & = & {\lambda}^2\left[ (|H_1|^2 + |H_2|^2)
|N|^2 + |H_1H_2|^2\right]
+ k^2|N|^4 - \lambda k (H_1H_2{N^{\ast}}^2 + H.c.) \nonumber \\
          & + & \frac{1}{8} ({g_1}^2 + {g_2}^2)(|H_2|^2 - |H_1|^2)^2
+ \frac{1}{2}{g_2}^2|{H_2}^{\dagger}H_1|^2 \nonumber \\
          & + & {m_{H_1}}^2|H_1|^2 + {m_{H_2}}^2|H_2|^2 +
{m_N}^2|N|^2 \nonumber \\
          & - &  \lambda A_{\lambda}(H_1H_2N + H.c.)
- \frac{1}{3} kA_k(N^3 + H.c.),
\end{eqnarray}
where $H_1H_2={H_1}^Ti{\sigma}_2H_2$,
${\sigma}_2$ is the second Pauli matrix and
${H_1}^T=({H_1}^0 \ \ {H_1}^-)$,
${H_2}^T=({H_2}^+ \ \ {H_2}^0)$.
The parameters $m_{H_i}$, $m_N$, $A_{\lambda ,k}$ are associated with
soft SUSY breaking terms.\footnote{In our analysis we shall not
consider any other soft SUSY breaking parameters.}
We shall take
the parameters $\lambda$, $k$, $A_{\lambda}$ and $A_k$
to be real and positive, which is a sufficient condition
for the vacuum to conserve CP and leads to a choice of
vacuum in which all the VEVs $x,v_1,v_2$ are real and positive \cite{4}.
Note that the range of the parameters is restricted by the condition
that the vacuum does not break QED,
which is not automatic in the NMSSM, and
is equivalent to the condition that ${m_c}^2 \geq 0$,
where $m_c$ is the mass of the physical charged Higgs $H^{\pm}$.
Also it is not automatic that the vacuum does not break QCD
although it has been checked that this is the case
for values of $r$ and $\tan \beta$ in the range 0.05-20 \cite{4}.
Finally, the range of the parameter $A_k$ is restricted by the
requirements that $<V_{Higgs}> < 0$, and all mass squared eigenvalues
are positive. We shall use the above conditions
in our analysis.

The effective theory below $M_{SUSY}$ is just the standard model
with two light Higgs doublets and a light Higgs singlet.
Thus the Higgs potential at some low energy scale $\mu <M_{SUSY}$
is given by the general expression
\footnote{There is a global $U(1)$ symmetry
of the quartic potential which
forbids the radiative generation of other terms not included in Eq.(4).}
\begin{eqnarray}
V_{Higgs} & = & \frac{1}{2} {\lambda}_1({H_1}^{\dagger}H_1)^2
+ \frac{1}{2} {\lambda}_2({H_2}^{\dagger}H_2)^2
+ ({\lambda}_3 +{\lambda}_4)({H_1}^{\dagger}H_1)({H_2}^{\dagger}H_2)
\nonumber \\
          & - & {\lambda}_4|{H_2}^{\dagger}H_1|^2
+ {\lambda}_5|N|^2|H_1|^2 + {\lambda}_6|N|^2|H_2|^2 \nonumber \\
          & + & {\lambda}_7({N^{\ast}}^2H_1H_2 + H.c.)
+ {\lambda}_8|N|^4 \nonumber \\
          & + & {m_1}^2|H_1|^2 + {m_2}^2|H_2|^2
           + {m_3}^2|N|^2 \nonumber \\
          & - &  m_4(H_1H_2N + H.c.)- \frac{1}{3} m_5(N^3 + H.c.).
\end{eqnarray}
Comparing Eqs.(3) and (4), the running quartic couplings ${\lambda}_i$
and the mass parameters $m_i$ must satisfy the following boundary
conditions at $M_{SUSY}$
\begin{eqnarray}
{\lambda}_1 &=& {\lambda}_2=\frac{1}{4} ({g_2}^2 + {g_1}^2),\ \ \
{\lambda}_3=\frac{1}{4} ({g_2}^2 - {g_1}^2) \nonumber \\
{\lambda}_4 &=& {\lambda}^2-\frac{1}{2}{g_2}^2,\ \ \
{\lambda}_5={\lambda}_6={\lambda}^2, \ \ \
{\lambda}_7=-{\lambda}k, \ \ \ {\lambda}_8=k^2, \nonumber \\
        m_1 & = & m_{H_1}, \ \ \ m_2  =  m_{H_2}, \ \ \
m_3  =  m_N, \nonumber \\
        m_4 &=& {\lambda}A_{\lambda}, \ \ \ m_5=kA_k.
\end{eqnarray}

At energy scales $\mu$ below $M_{SUSY}$, the values of the
quartic couplings may be obtained by solving the following
RG equations which we have derived for the potential in Eq.(4)
with the aid of the general results in ref.\cite{18}:
\begin{eqnarray}
16{\pi}^2\frac{\partial {\lambda}_1}{\partial t}
& = &
12{{\lambda}_1}^2 \ \ + \ \ 4{{\lambda}_3}^2
\ \  + \ \ 4{\lambda}_3{\lambda}_4
\ \ + \ \ 2{{\lambda}_4}^2\ \  + \ \ 2{{\lambda}_5}^2    \nonumber \\
& - &
{\lambda}_1(3{g_1}^2 + 9{g_2}^2)
\ \ + \ \ \frac{3}{4} {g_1}^4
\ \ + \ \ \frac{9}{4} {g_2}^4\ \  + \ \ \frac{3}{2} {g_1}^2{g_2}^2,
\nonumber \\
16{\pi}^2\frac{\partial {\lambda}_2}{\partial t}
& = &
12{{\lambda}_2}^2 \ \ + \ \ 4{{\lambda}_3}^2\ \  + \ \
4{\lambda}_3{\lambda}_4
\ \ + \ \ 2{{\lambda}_4}^2\ \  + \ \ 2{{\lambda}_6}^2    \nonumber \\
& - &
{\lambda}_2(3{g_1}^2 + 9{g_2}^2)
\ \ + \ \ \frac{3}{4} {g_1}^4
\ \ + \ \ \frac{9}{4} {g_2}^4\ \  + \ \ \frac{3}{2} {g_1}^2{g_2}^2
\nonumber \\
& + &
12{h_t}^2{\lambda}_2\ \ - \ \ 12{h_t}^4,
\nonumber \\
16{\pi}^2\frac{\partial {\lambda}_3}{\partial t}
& = &
2({\lambda}_1 + {\lambda}_2)(3{\lambda}_3 + {\lambda}_4) \ \ +
\ \ 4{{\lambda}_3}^2\ \  + \ \ 2{{\lambda}_4}^2
\ \ 2{\lambda}_5{\lambda}_6\ \   \nonumber \\
& - &
{\lambda}_3(3{g_1}^2 + 9{g_2}^2)
\ \ + \ \ \frac{3}{4} {g_1}^4
\ \ + \ \ \frac{9}{4} {g_2}^4\ \  - \ \ \frac{3}{2} {g_1}^2{g_2}^2
\ \ + \ \ 6{h_t}^2{\lambda}_3,
\nonumber \\
16{\pi}^2\frac{\partial {\lambda}_4}{\partial t}
& = &
2{\lambda}_4({\lambda}_1 + {\lambda}_2 + 4{\lambda}_3 + 2{\lambda}_4)
\ \ + \ \ 4{{\lambda}_7}^2     \nonumber \\
& - &
{\lambda}_4(3{g_1}^2 + 9{g_2}^2)\ \ + \ \ 3{g_1}^2{g_2}^2
\ \ + \ \ 6{h_t}^2{\lambda}_4,
\nonumber \\
16{\pi}^2\frac{\partial {\lambda}_5}{\partial t}
& = &
2{\lambda}_5(3{\lambda}_1 + 2{\lambda}_5 + 4{\lambda}_8)
\ \ + \ \  2{\lambda}_6(2{\lambda}_3 + {\lambda}_4)
\ \ + \ \ 8{{\lambda}_7}^2     \nonumber \\
& - &
\frac{1}{2} {\lambda}_5(3{g_1}^2 + 9{g_2}^2),
\nonumber \\
16{\pi}^2\frac{\partial {\lambda}_6}{\partial t}
& = &
2{\lambda}_5(2{\lambda}_3 + {\lambda}_4)
\ \ + \ \  2{\lambda}_6(3{\lambda}_2 + 2{\lambda}_6 + 4{\lambda}_8)
\ \ + \ \ 8{{\lambda}_7}^2     \nonumber \\
& - &
\frac{1}{2} {\lambda}_6(3{g_1}^2 + 9{g_2}^2)
\ \ + \ \ 6{h_t}^2{\lambda}_6,
\nonumber \\
16{\pi}^2\frac{\partial {\lambda}_7}{\partial t}
& = &
2{\lambda}_7({\lambda}_3 + 2{\lambda}_4 + 2{\lambda}_5 + 2{\lambda}_6
+ 2{\lambda}_8)     \nonumber \\
& - &
\frac{1}{2} {\lambda}_7(3{g_1}^2 + 9{g_2}^2)
\ \ + \ \ 3{h_t}^2{\lambda}_7,
\nonumber \\
16{\pi}^2\frac{\partial {\lambda}_8}{\partial t}
& = &
2{{\lambda}_5}^2 + 2{{\lambda}_6}^2 + 4{{\lambda}_7}^2 +
20{{\lambda}_8}^2,
\end{eqnarray}
where $t=\ln \mu$, and $h_t$ obeys the RG equation
\begin{eqnarray}
16{\pi}^2\frac{\partial h_t} {\partial t}
& = &
\frac{9}{2} {h_t}^3
\ \ - \ \ h_t(8{g_3}^2 + \frac{9}{4} {g_2}^2 + \frac{17}{12} {g_1}^2).
\end{eqnarray}
The gauge couplings $g_i$ obey the RG equations of the standard
model with two Higgs doublets and three fermion families
below $M_{SUSY}$, namely
\begin{equation}
16{\pi}^2\frac{\partial g_i} {\partial t}=-c_i{g_i}^3
\ \ \ (c_1=-7,c_2=3,c_3=7),
\end{equation}
and we shall take $\alpha _{1,2,3}(M_Z)=0.0102,0.0336,0.113$,
respectively.

The minimisation conditions implied by
$\frac{ \partial V_{Higgs}}{\partial v_i}=0$ and
$\frac{ \partial V_{Higgs}}{\partial x}=0$
allow us to eliminate the low energy parameters
$m_1$, $m_2$, $m_3$. The remaining masses $m_4$ and $m_5$
are related to the parameters $A_{\lambda}$ and $A_k$ at $M_{SUSY}$
by Eq.(5). Below this scale we shall regard $m_4$ and $m_5$ as
free parameters.
The charged Higgs squared mass is given by
\begin{eqnarray}
{m_c}^2& = & \frac{2x}{\sin 2\beta}(m_4 - {\lambda}_7x)
\ \ - \ \ {\lambda}_4v^2,
\end{eqnarray}
which allows us to eliminate the parameter $m_4$ in favour
of the more physical parameter ${m_c}^2$.
The physical neutral CP-odd pseudoscalar mass squared
symmetric matrix elements are
\begin{eqnarray}
(M_{ps}^2)_{11} & = & 3xm_5 + \frac{v^2}{4x^2}\sin ^22\beta
(m_c^2 + \lambda _4v^2) -\frac{3}{2}\lambda _7v^2\sin 2\beta,
\nonumber \\
(M_{ps}^2)_{12} & = &
\frac{v}{2x}\sin 2\beta (m_c^2 + \lambda _4 v^2)
+ 3vx\lambda _7, \nonumber \\
(M_{ps}^2)_{22} & = & m_c^2 + \lambda _4 v^2.
\end{eqnarray}
The neutral CP-even scalar mass squared symmetric matrix elements are,
in the basis $1,2,3=H_1,H_2,N$,
\begin{eqnarray}
(M_{s}^2)_{11} & = &
2{\lambda}_1{v_1}^2 + (m_c^2 + \lambda _4v^2) \sin ^2 \beta ,
\nonumber \\
(M_{s}^2)_{12} & = &
\sin 2\beta[({\lambda}_3 + \frac{1}{2}{\lambda}_4)v^2
- \frac{1}{2}m_c^2] ,\nonumber \\
(M_{s}^2)_{22} & = &
2{\lambda}_2{v_2}^2 +
\cos ^2 \beta( m_c^2 + \lambda _4v^2), \nonumber \\
(M_{s}^2)_{13} & = &
x(2{\lambda}_5v_1 + {\lambda}_7v_2)
-\frac{v_2}{2x}\sin 2\beta (m_c^2 + \lambda _4v^2), \nonumber \\
(M_{s}^2)_{23} & = &
x(2{\lambda}_6v_2 + {\lambda}_7v_1)
- \frac{v_1}{2x}\sin 2\beta (m_c^2 + \lambda _4v^2), \nonumber \\
(M_{s}^2)_{33} & = &
4\lambda _8x^2 + \frac{v^2}{4x^2}\sin ^22\beta
(m_c^2 + \lambda _4v^2) +\frac{1}{2}\lambda _7v^2\sin 2\beta
-xm_5.
\end{eqnarray}

The model is then specified by the 6 parameters $\lambda$, $k$,
$\tan \beta=v_2/v_1$, $r=x/v$, $m_5$, and ${m_c}^2$.
Our detailed procedure for evaluating Higgs masses is as follows.
Given the gauge couplings $g_i(M_Z)$,
we find $g_i(m_t)$ by solving the RG equations (8).
For a given $m_t$
and $\tan \beta$ we obtain $h_t(m_t)$,
then use its RG equation (7)
and those for the gauge couplings (8)
to evaluate $h_t(M_{SUSY})$ and $g_i(M_{SUSY})$.
Given these couplings and some choice
of $\lambda$ and $k$ we then use the boundary conditions
in Eq.(5) and the RG equations (6),(7),(8),
to find ${\lambda}_i(\mu)$,
where $\mu$ is close to mass-shell for the Higgs boson of interest.
\footnote{In integrating the RG equations from $M_{SUSY}$
down to $\mu$ we have decoupled the terms involving $h_t$
below the scale $m_t$, but we have not decoupled any
Higgs bosons below their mass scales. We would not expect
such a decoupling to change our results significantly
providing there are no very heavy scalars.}
In practice we shall take $M_{SUSY}=1$ TeV, $\mu =150$ GeV.
The neutral Higgs masses are then calculated from
Eqs.(10),(11) with ${\lambda}_i={\lambda}_i(\mu)$.
The diagonalisation of these matrices also gives the mixing angles,
so that the amplitude of the singlet state $N$ in each of the neutral
mass eigenstates may be calculated.

We shall now obtain an upper bound
on the mass of the lightest neutral scalar Higgs boson $h^0$,
by using the fact that $m_h^2$ must not exceed
the lower eigenvalue of the upper $2 \times 2$ block matrix,
$(M_s^2)_{ij}$, $i,j=1,2$.
The resulting upper
bound is a complicated function of ${m_c}^2$
\begin{equation}
{m_h}^2\leq \frac{1}{2} (A + {m_c}^2)
\ \ - \ \ \frac{1}{2} \sqrt{({m_c}^2 + B )^2
+ C^2 - B^2},
\end{equation}
where
\begin{eqnarray}
A & = & v^2({\lambda}_1 + {\lambda}_2 + {\lambda}_4)
+ v^2({\lambda}_1 - {\lambda}_2)\cos 2\beta ,   \nonumber \\
B & = & -v^2\cos 2\beta[({\lambda}_1 - {\lambda}_2)
+ ({\lambda}_1 +{\lambda}_2 - {\lambda}_4)\cos 2\beta]
- 2v^2\sin ^22\beta({\lambda}_3 + \frac{1}{2}{\lambda}_4), \nonumber \\
C^2 & = & v^4[({\lambda}_1 - {\lambda}_2)
+({\lambda}_1 + {\lambda}_2 - {\lambda}_4)\cos 2\beta ]^2
+ 4v^4\sin ^22\beta ({\lambda}_3 + \frac{1}{2} {\lambda}_4)^2.
\end{eqnarray}
It is easy to show that the above bound reaches a maximum
asymptotically for $m_c\rightarrow \infty$.
Since $(C^2 - B^2)\geq 0$,
we obtain the $m_c$-independent bound
\begin{equation}
{m_h}^2\leq \frac{1}{2} (A-B).
\end{equation}
Inserting $A,B$ from Eq.(13) and using the boundary conditions
in Eq.(5) leads to an upper bound of the form
\begin{eqnarray}
{m_{h}}^2 & \leq & {M_Z}^2 +
(\lambda^2v^2-{M_Z}^2)\sin^22\beta \nonumber \\
& + & \frac{v^2}{2}[(\delta {\lambda}_1 + \delta {\lambda}_2 +
2\delta {\lambda}_3 + 2\delta {\lambda}_4)
+ 2(\delta {\lambda}_1 - \delta {\lambda}_2)\cos 2\beta \nonumber \\
& + & (\delta {\lambda}_1 + \delta {\lambda}_2
- 2\delta {\lambda}_3 - 2\delta {\lambda}_4)\cos ^22\beta],
\end{eqnarray}
where $\delta {\lambda}_i= {\lambda}_i(\mu ) - {\lambda}_i(M_{SUSY})$,
and, strictly speaking,
$M_Z$ should be evaluated at $M_{SUSY}$.
Using the RG equations (6), keeping only terms
quadrilinear in $\lambda$, $h_t$, and $k$
and making a small $\delta {\lambda}_i$ approximation,
Eq.(15) yields the analytic bound
\footnote{Note that Eq.(16) reduces to the bound in the MSSM in the limit
that $\lambda \rightarrow 0$.}
\begin{eqnarray}
{m_{h}}^2 & \leq & {M_Z}^2 +
(\lambda^2v^2-{M_Z}^2)\sin^22\beta \nonumber \\
& + & \frac{v^2}{32\pi ^2}\ln \left( \frac{M_{SUSY}}{\mu} \right)
[(12{h_t}^4 - 12{\lambda}^2{h_t}^2 - 8{\lambda}^2k^2 - 24{\lambda}^4)
-24{h_t}^4\cos 2\beta \nonumber \\
& + & (12{h_t}^4 + 12{\lambda}^2{h_t}^2 + 8{\lambda}^2k^2 + 8{\lambda}^4)
\cos ^22\beta ].
\end{eqnarray}
In Eq.(16), the tree level bound in Eq.(2) is corrected by a function of
$h_t$,  $\cos 2\beta$, $\lambda$ and $k$.
It is easy to show that this function is maximised
for $\lambda >{\lambda}_{max}$ and $k=0$,
and so we shall use $\lambda _{max}$ and $k=0$,
as in the case of the tree level bound discussed previously.
With this information in hand,
let us now return to the more exact result in Eq.(15)
in which the values of $\delta {\lambda}_i$ are obtained
from numerically integrating the RG equations,
decoupling the top quark below its mass and choosing $\mu=150$ GeV.
For each value of $m_t$, we have determined the value of $h_t(M_{SUSY})$
and the corresponding value of ${\lambda}_{max}$ which
maximise the r.h.s.
of Eq.(15) from a numerical analysis of
the triviality condition as discussed previously.
These are shown in Fig.1, together with $\sin \beta$
which is eliminated from Eq.(15) using
$m_t=h_t(m_t)v\sin \beta$.

Using the parameters in Fig.1, we obtain the
upper bound on the lightest neutral Higgs scalar in the NMSSM
as a function of $m_t$ as shown in Fig.2.
We again emphasise that we have considered the more general case of two
light Higgs doublets plus a light singlet all below the
scale $M_{SUSY}$, whereas previous results \cite{16,17}
have only considered the case of one light Higgs boson.
{}From Fig.2 we find that for $m_t=90$ GeV, $m_h\leq 145$ GeV,
which agrees to within 2 GeV with both refs.\cite{16,17}.
Our bound for larger top
quark masses, although qualitatively similar to refs.\cite{16,17},
can differ by several GeV.
It can easily be checked that
the analytic formula in Eq.(16) agrees with the full numerical
bound shown to within 5 GeV across the entire $m_t$ region,
giving a bound somewhat lower than the true bound for $m_t=90-160$
GeV, somewhat higher for $m_t=180-190$ GeV, and in good agreement
for $m_t\approx 170$ GeV. If the $\lambda$ dependent
terms in the radiative corrections in Eq.(16) were dropped, this
would produce a bound about 7 GeV higher
than the true bound over the entire $m_t$ range.

We shall now discuss the physical parameter regions in which
the upper bound on $m_h$ in Fig.2 is approximately realised,
and discuss the light Higgs boson spectrum in these regions.
As mentioned earlier, the $m_c$ independent bound
in Eqs.(15),(16) is exactly valid only asymptotically, but actually
the bound rapidly approaches its maximum for large values
of $m_c$ of the order of the bound itself.
Since it is known that the allowed range of $m_c$ is controlled
by $r$ and that larger values of $m_c$ are associated with the
larger values of $r$, we shall restrict ourselves
to $r=1.0$ and $10.0$, as shown in Figs.3 and 4.

Fig.3(a) shows the light Higgs boson spectrum as a function
of $m_c$ for $m_t=150$ GeV,
for $r=1.0$, where $\tan \beta=1.7$ is from Fig. 1.
We have taken $\lambda =0.65$ and $k=0.1$,
rather than ${\lambda}_{max}=0.7$ and $k=0$
since these values would imply a massless CP-odd scalar,
which is phenomenologically ruled out.
We have also taken $m_5=0$.
Only the lightest neutral CP-even scalars (solid)
and the lightest CP-odd pseudoscalar (dashed) are
included in Fig.3(a). The lightest scalar mass $m_h$
reaches a maximum of $m_h=88$ GeV.
In Fig.3(b) we show the $N$ components of the bosons in Fig.3(a).
This is of interest since the $N$ component is a gauge singlet, and thus
decoupled from the gauge bosons. The lightest CP-even scalar
is about 80\% decoupled rising to being almost 100\% decoupled
at its maximum.
It is obvious that the bound which follows from the upper
$2\times 2$ block matrix can only approach saturation
if the full scalar matrix is approximately block diagonal.
Since the lightest scalar is approximately decoupled near
its maximum mass, the physically observable second lightest scalar
must respect the bound at this point.
Taking $m_5$ to be non-zero and positive
(negative) serves to lower (raise) $(M_s^2)_{33}$
(identified as the decoupled scalar mass)
while leaving the other elements unchanged,
as is clear from Eq.(11).
Finally it is clear from Figs.3(a),3(b) that the lightest CP-odd state
presents no phenomenological problems,
since as well as being heavy it is over 90\% decoupled over
the entire $m_c$ range.

In Fig.4(a) we show the light Higgs spectrum for $r=10.0$,
choosing the other parameters to be the same as in Fig.3(a).
The relevant range of $m_c$ is now 2.3-2.5 TeV which is
strictly outside the range of validity of our calculation,
since we do not consider Higgs decoupling and choose $\mu=150$ GeV
(of order
$m_h$) as in Fig.3(a).
Nevertheless Fig.4(a) shows that the lightest CP-even scalar mass
reaches a maximum
of 118 GeV, close to the minimum of the physically observable second
lightest
scalar in Fig.3(a).
Fig.4(b) shows that
in this case the lightest scalar
is the detectable one, being almost 0\% decoupled at its maximum,
while the second lightest scalar is almost 100\% decoupled at this point.
The CP-odd boson in Fig.4 is now nearly 100\% decoupled over
the whole range of $m_c$.

In conclusion, the lightest CP-even neutral Higgs boson
in the NMSSM must respect the bounds,
$m_h< 145$ GeV for $m_t=90$ GeV,
decreasing to $m_h< 123$ GeV for $m_t=180$ GeV.
Our analysis is based on the assumption of a SUSY desert
between $M_{SUSY}=1$ TeV and $M_{GUT}=10^{16}$ GeV,
and involves the assumption of hard decoupling of the
superpartners. However, our calculations do
include the effects of other light Higgs
bosons which previous analyses have ignored.
We have presented an analytic expression in Eq.(16) for the
radiative corrections in the NMSSM, which agrees well with our
full numerical bound in Fig.2.
In addition we have discussed the light Higgs boson spectrum
in a region of parameter space which approximately realises the
bound, and have seen that there are indeed other light Higgs
bosons whose effects must be taken into account.
For $r\sim 1$ (Fig.3), the lightest CP-even scalar is not very
strongly physically coupled when it is close to its maximum mass,
and the second lightest CP-even
scalar, which must also respect the bound, may be easier to discover.

{\bf Acknowledgements}

TE and SFK would like to thank the SERC for financial support.

\clearpage

\section*{Figure Captions}

\noindent

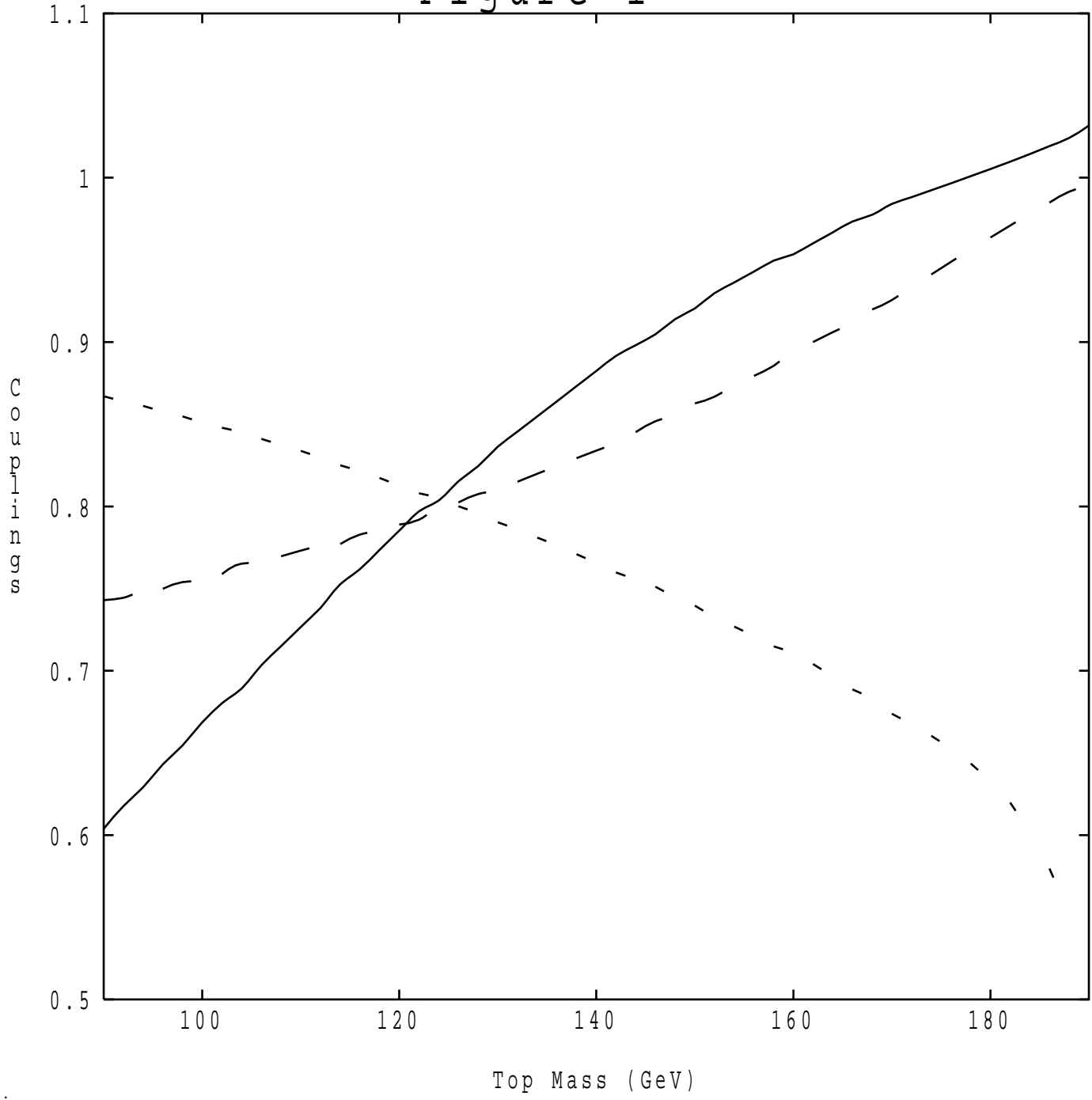
{\bf Figure 1} : The values of $h_t$ (solid), $\lambda_{max}$
(short dashed) and $\sin \beta$ (long dashed)
realising the upper bound on $m_h$ in Figure 2.

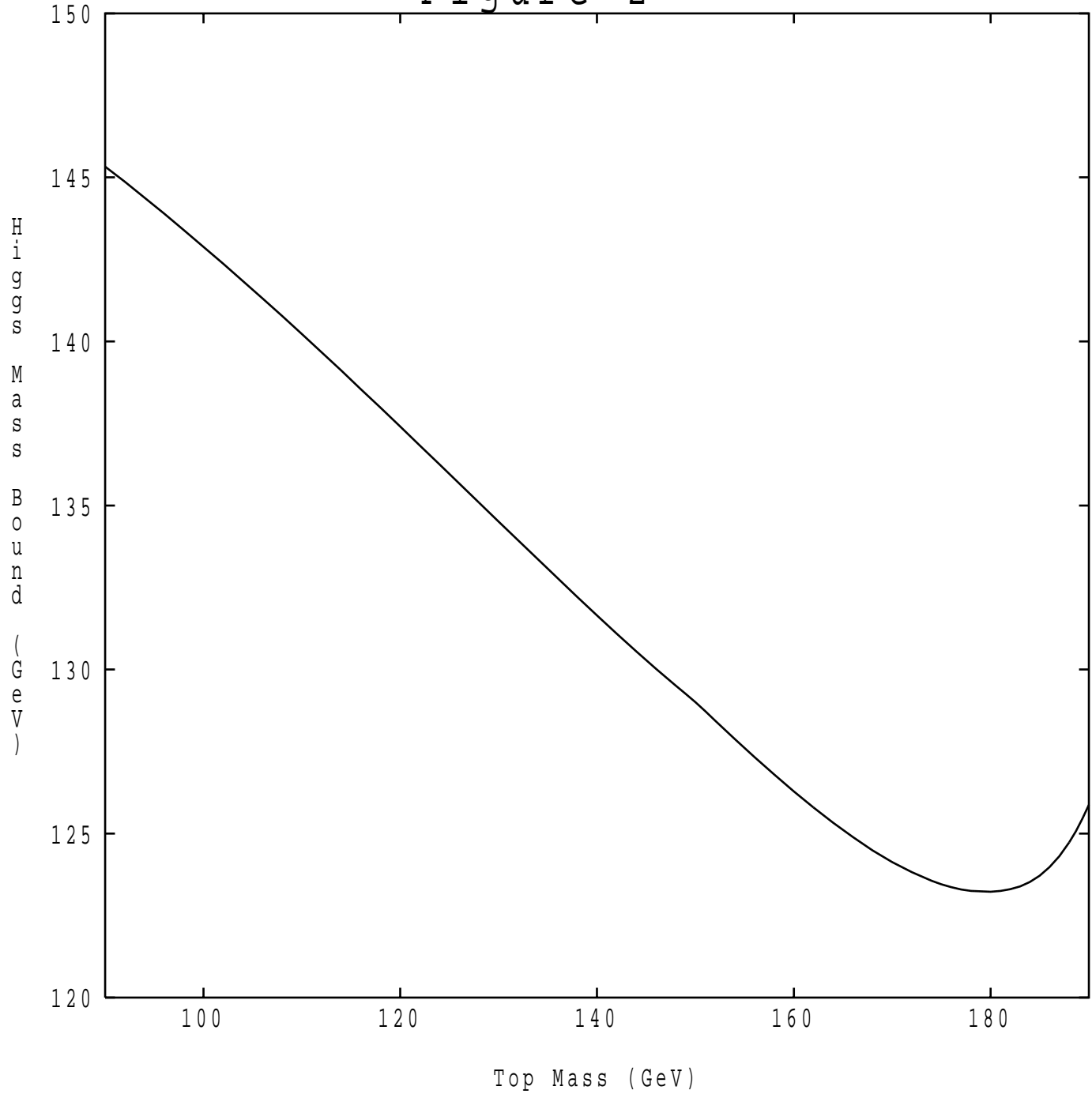
{\bf Figure 2} : The full numerical bound on the lightest neutral
CP-even Higgs scalar as a function of $m_t$.

{\bf Figure 3(a)} : The light Higgs boson spectrum for the parameters
$m_t$=150 GeV, $r=1.0$, $\tan \beta =1.7$, $\lambda =0.65$,
$k=0.1$. The solid lines are CP-even scalar masses,
and the dashed line is the CP-odd mass.

{\bf Figure 3(b)} : The $N$ components of each of the bosons
in Fig.3(a) where the solid line is for the lightest CP-even scalar,
the dashed line is for the second lightest CP-even scalar,
and the dotted line is for the CP-odd pseudoscalar.

{\bf Figure 4(a)} : The light Higgs boson spectrum for $r=10.0$.
The other parameters are as in Figure 3(a).
The solid lines are CP-even scalar masses,
and the dashed line is the CP-odd mass.

{\bf Figure 4(b)} : The $N$ components of each of the bosons
in Fig.4(a) where, as in Fig.3(b),
the solid line is for the lightest CP-even scalar,
the dashed line is for the second lightest CP-even scalar,
and the dotted line is for the CP-odd pseudoscalar.

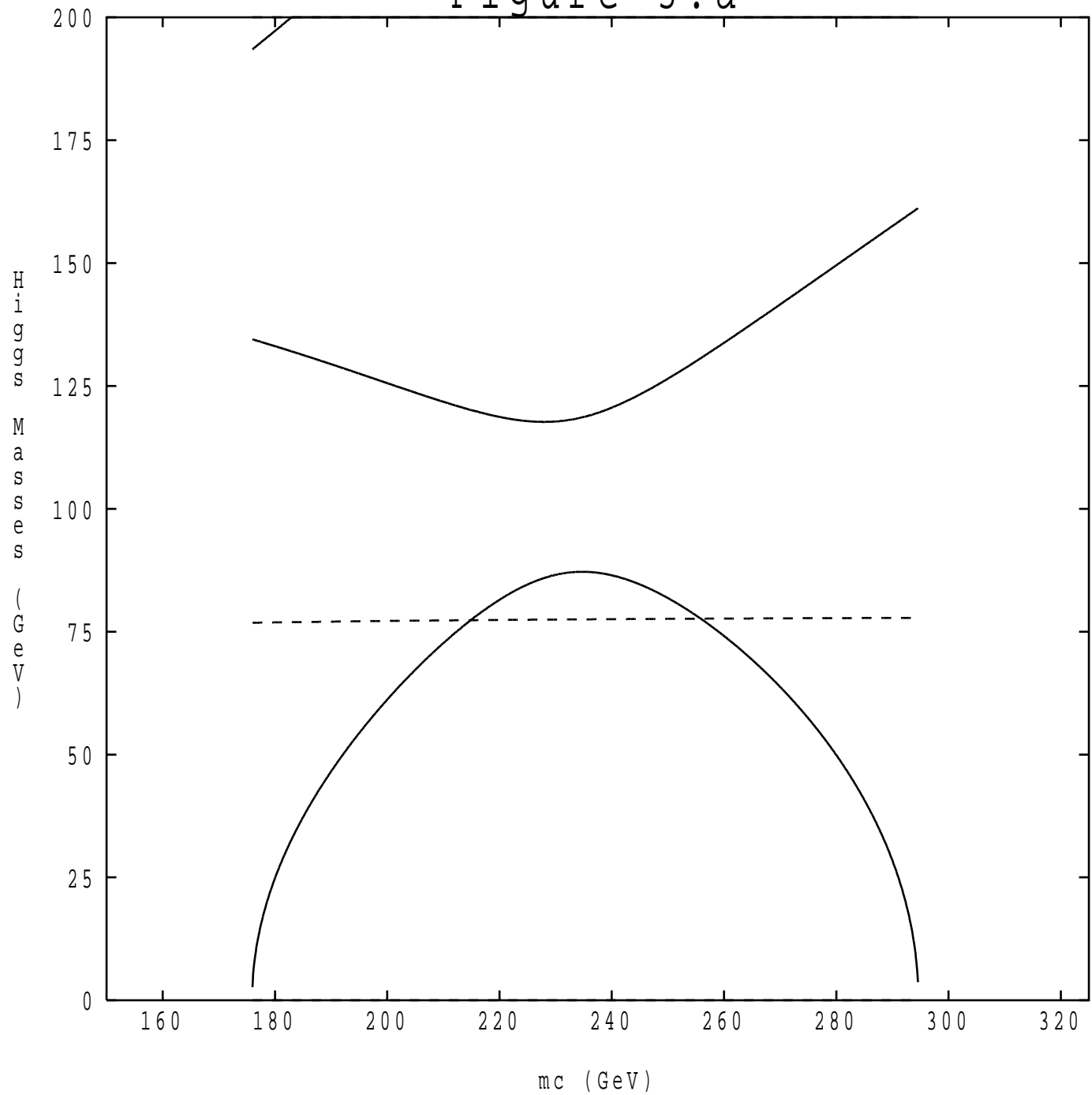
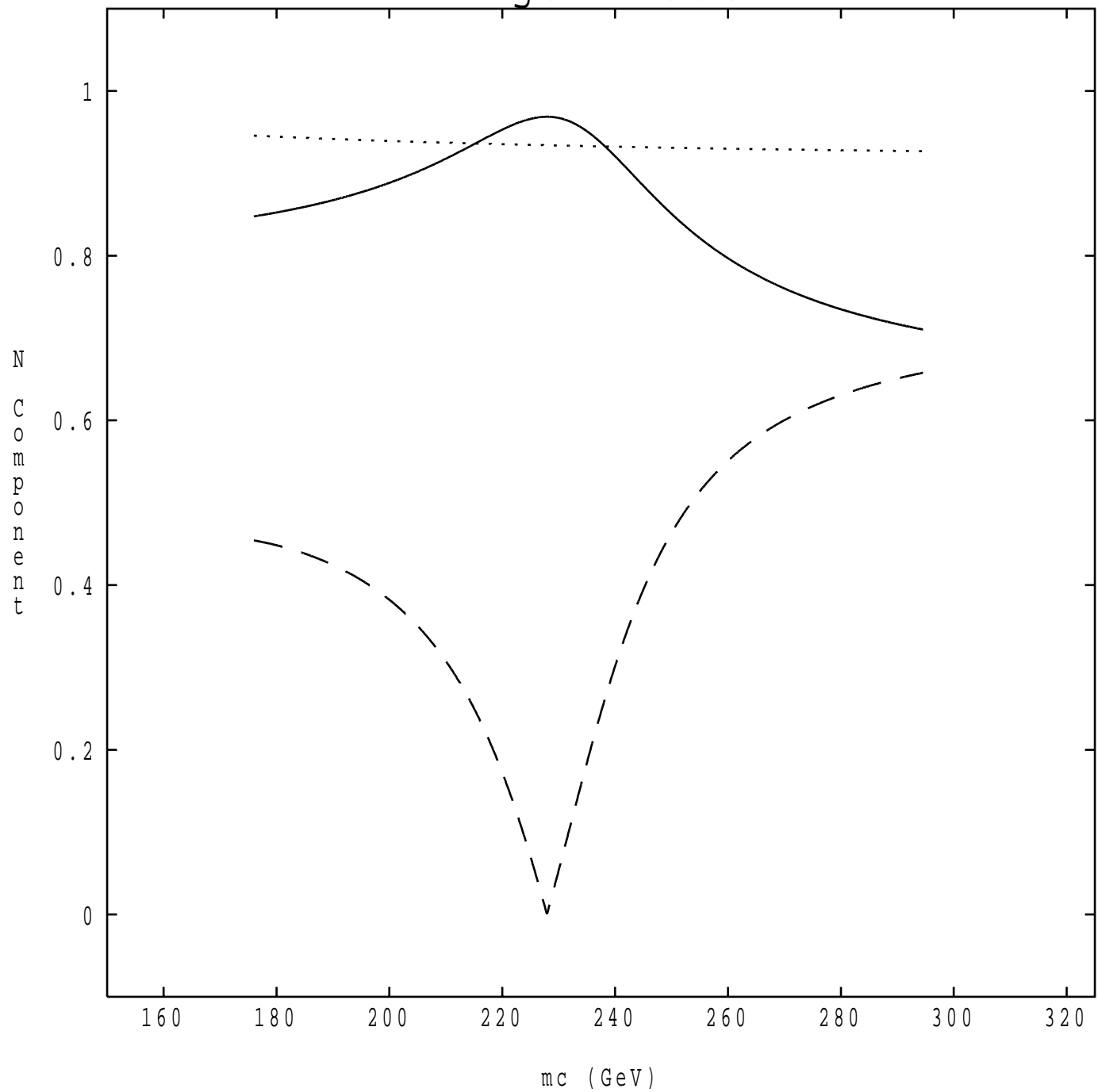
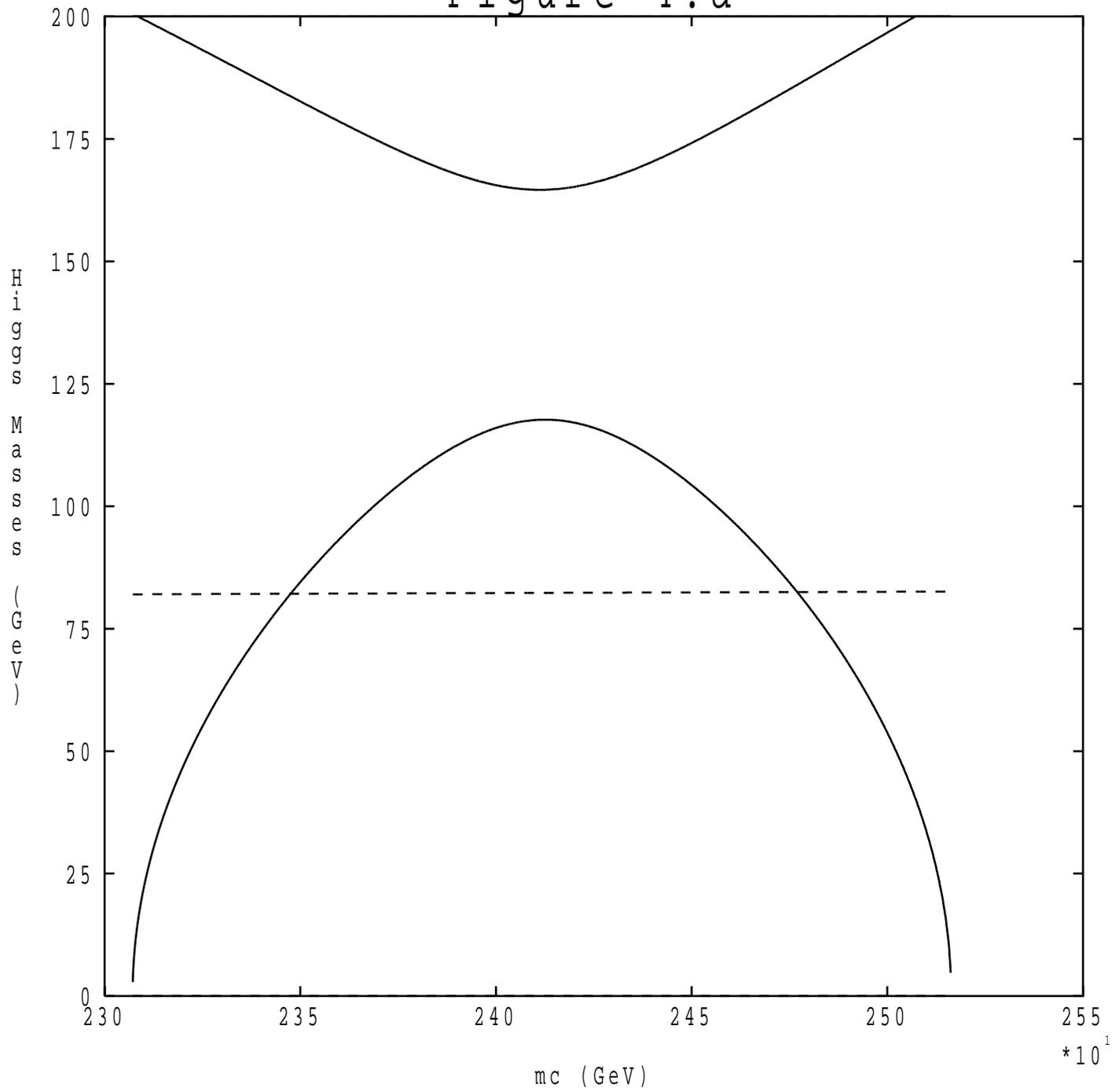
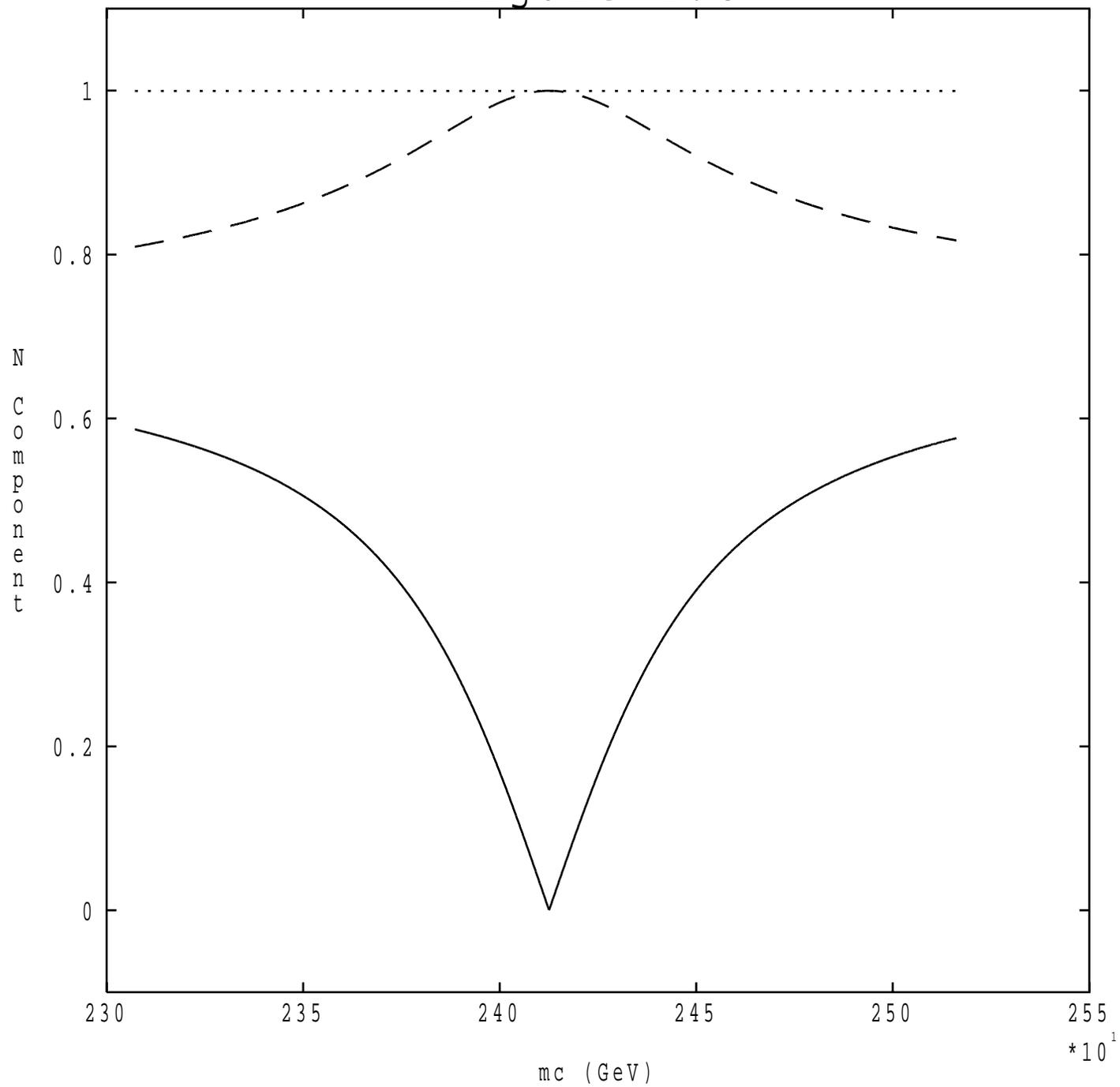
\end{document}